\pdfoutput=1
\newif\ifdraft
\newif\iffull
\newif\ifcomment
\newif\iflatexdiff
\newif\ifbibtex
\newif\ifpreprint
\preprinttrue
\def\dvers{v1.8}
\ifpreprint
\documentclass[ALICE,manyauthors]{cernphprep}
\usepackage[comma,square,numbers,sort&compress]{natbib}
\usepackage{hyperref}
\else
\documentclass[10pt,aps,prl,twocolumn,superscriptaddress,altaffilletter,nobibnotes,nofootinbib]{revtex4-1}
\fi
\usepackage{graphicx}  
\usepackage{dcolumn}   
\usepackage{amssymb}   
\usepackage{amsfonts}
\usepackage{graphics}
\usepackage{epsfig}
\usepackage[usenames]{color}
\iflatexdiff
\RequirePackage[normalem]{ulem}
\RequirePackage{color}\definecolor{RED}{rgb}{1,0,0}\definecolor{BLUE}{rgb}{0,0,1}

\fi
\ifpreprint
\else

\fi

\newcommand{\ZDC}          {\rm{ZDC}}
\newcommand{\ZDCs}         {\rm{ZDCs}}
\newcommand{\SPD}          {\rm{SPD}}

\newcommand{\VZERO}        {\rm{VZERO}}
\newcommand{\VZEROA}       {\rm{VZERO-A}}
\newcommand{\VZEROC}       {\rm{VZERO-C}}

\newcommand{\pp}           {pp}

\newcommand{\PbPb}         {\mbox{Pb--Pb}}
\newcommand{\AuAu}         {\mbox{Au--Au}}

\newcommand{\dNdeta}       {\mathrm{d}N_\mathrm{ch}/\mathrm{d}\eta}

\newcommand{\snn}          {\ensuremath{\sqrt{s_{\rm NN}}}}
\newcommand{\snnbf}        {\ensuremath{\mathbf{{\sqrt{s_{\mathbf NN}}}}}}

\newcommand{\Npart}        {\ensuremath{N_\mathrm{part}}}
\newcommand{\avNpart}      {\ensuremath{\langle N_\mathrm{part} \rangle}}

\newcommand{\Ncoll}        {\ensuremath{N_\mathrm{coll}}}

\newcommand{\dNdetape}     {\left(\ensuremath{\dNdeta\right)/\left(\avNpart/2\right)}}

\newcommand{\abs}[1]       {\ensuremath{\left|#1\right|}}

\newcommand{\Fig}[1]       {Fig.~\ref{#1}}
\newcommand{\Figure}[1]    {Figure~\ref{#1}}

\newcommand{\warn}[1]      {{\small\textbf{(!\footnote{\textbf{(!)}~#1})}}\marginpar{\textbf{---}}}

\newcommand{\final}[1]     {\textbf{\textcolor{blue}{#1}}}

\ifdraft
\usepackage{lineno}
\linenumbers
\modulolinenumbers[5]
\usepackage{fancyhdr}
\renewcommand{\final}[1]{#1}
\else
\renewcommand{\warn}[1]{}
\renewcommand{\final}[1]{#1}
\fi
\begin{document}
\newlength{\figlen}
\setlength{\figlen}{\linewidth}
\ifpreprint
\setlength{\figlen}{0.75\textwidth}
\begin{titlepage}
\PHnumber{2010-071}                   
\PHdate{06 Dec 2010}                  
\title{Centrality dependence of the charged-particle multiplicity density\\ 
at mid-rapidity in \PbPb\ collisions at \snnbf\ = 2.76 TeV}
\ShortTitle{Centrality dependence of the charged-particle multiplicity}
\Collaboration{ALICE Collaboration%
         \thanks{See Appendix~\ref{app:collab} for the list of collaboration members}}
\ShortAuthor{ALICE Collaboration} 
\else
\title{Centrality dependence of the charged-particle multiplicity density\\ 
at mid-rapidity in \PbPb\ collisions at \snnbf\ = 2.76 TeV}
\iffull
\input{authors-prl.tex}
\else
\collaboration{The ALICE Collaboration}
\fi
\vspace{0.3cm}
\ifdraft
\date{\today, \color{red}DRAFT \dvers\ \$Revision: 440 $\color{white}:$\$\color{black}}
\else
\date{\today}
\fi
\fi
\begin{abstract}
The centrality dependence of the charged-particle multiplicity density at mid-rapidity
in \PbPb\ collisions at $\snn= 2.76$ TeV is presented. 
The charged-particle density normalized per participating nucleon pair increases by about a factor of
two from peripheral~(70--80\%) to central~(0--5\%) collisions. 
The centrality dependence is found to be similar to that observed at lower collision energies.
The data are compared with models based on different mechanisms for particle production in nuclear
collisions.
\end{abstract}
\ifpreprint
\end{titlepage}
\setcounter{page}{2}
\else
\maketitle
\fi
\ifdraft
\thispagestyle{fancyplain}
\fi
%
Quantum Chromodynamics~(QCD), the theory of the strong interaction, predicts a phase transition at high temperature
between hadronic and deconfined matter~(the Quark--Gluon Plasma).
Strongly interacting matter under such extreme conditions can be studied experimentally using 
ultra-relativistic collisions of heavy nuclei. 
The field entered a new era in November 2010 when the Large Hadron Collider~(LHC) at CERN produced the first 
\PbPb\ collisions at a centre-of-mass energy per nucleon pair $\snn = 2.76$ TeV. 
This represents an increase of more than one order of magnitude over the highest energy nuclear collisions previously 
obtained in the laboratory.

The multiplicity of charged particles produced in the central rapidity region is a key observable 
to characterize the properties of the matter created in these collisions~\cite{Armesto:2009ug}.
Nuclei are extended objects, and their collisions can be characterized by centrality, related to the collision 
impact parameter.
The study of the dependence of the charged-particle density on colliding system, centre-of-mass
energy and collision geometry is important to understand the relative contributions to particle production 
of hard scattering and soft processes, and may provide insight into the partonic structure of the projectiles.

The ALICE Collaboration recently reported the measurement of the charged-particle pseudo-rapidity 
density at mid-rapidity for the most central~(head-on) \PbPb\ collisions at $\snn = 2.76$ TeV~\cite{Aamodt:2010pb}.
In this Letter, we extend that study to non-central collisions, presenting the measurement of the centrality 
dependence of the multiplicity density of charged primary particles $\dNdeta$ in the pseudo-rapidity interval 
$\abs{\eta}<0.5$. 
The pseudo-rapidity is defined as $\eta \equiv - \ln \tan (\theta/2)$ where $\theta$ is the angle between 
the charged-particle direction and the beam axis~($z$). 
Primary particles are defined as all prompt particles produced in the collision, including decay products, 
except those from weak decays of strange particles. 

We report the charged-particle density per participant-pair, $\dNdetape$, for nine centrality classes, covering 
the most central 80\% of the hadronic cross section. 
The average number of nucleons participating in the collision in a given centrality class, $\avNpart$,
reflects the collision geometry and is obtained using Glauber modeling~\cite{Miller:2007ri}.
The results are compared with measurements at lower collision energy~\cite{Adcox:2000sp,Back:2001xy,Back:2002uc,
Adler:2004zn,Abelev:2008ez,Alver:2010ck} 
and with theoretical calculations~\cite{Bopp:2007sa,Deng:2010xg,Armesto:2004ud,Kharzeev:2004if,ALbacete:2010ad}. 

The data for this measurement were collected with the ALICE detector~\cite{aliceapp}.
The data sample is the same as in~\cite{Aamodt:2010pb} and the analysis techniques are similar. 
The main detector utilized in the analysis is the Silicon Pixel Detector (\SPD), 
the innermost part of the Inner Tracking System~(ITS).
The SPD consists of two cylindrical layers of hybrid silicon pixel assemblies covering
$|\eta|<2.0$ and $|\eta|<1.4$ for the inner and outer layers, respectively.
A total of $9.8\times 10^{6}$ pixels of size $50\times425$~$\mu$m$^2$ are read out by 1200 electronic chips.
Each chip also provides a fast signal when at least one of its pixels is hit.
These signals are combined in a programmable logic unit which supplies a trigger signal. 
A trigger signal is also provided by the \VZERO\ counters, two arrays of 32 scintillator 
tiles covering the full azimuth within $2.8<\eta<5.1$~(\VZEROA) and $-3.7<\eta<-1.7$~(\VZEROC). 
The trigger was configured for high efficiency for hadronic events, requiring at least two out of the following 
three conditions: 
i) two pixel chips hit in the outer layer of the \SPD, 
ii) a signal in \VZEROA, 
iii) a signal in \VZEROC.
The threshold in the \VZERO\ detector corresponds approximately to the energy deposition of a 
minimum ionizing particle.
This trigger configuration led to a rate of about 50~Hz, with 4~Hz from nuclear interactions, 
45~Hz from electromagnetic processes, and 1~Hz arising from beam background.
In addition, in the offline event selection, we also use the information from two neutron Zero Degree 
Calorimeters~(\ZDCs) positioned at $\pm$~114~m from the interaction point. 
Beam background events are removed using the \VZERO\ and \ZDC\ timing information. 
Electromagnetically induced interactions are reduced by requiring an energy deposition above 500 GeV in 
each of the neutron \ZDCs.

\begin{figure}[tbh]
\begin{center}
\includegraphics[width=\figlen]{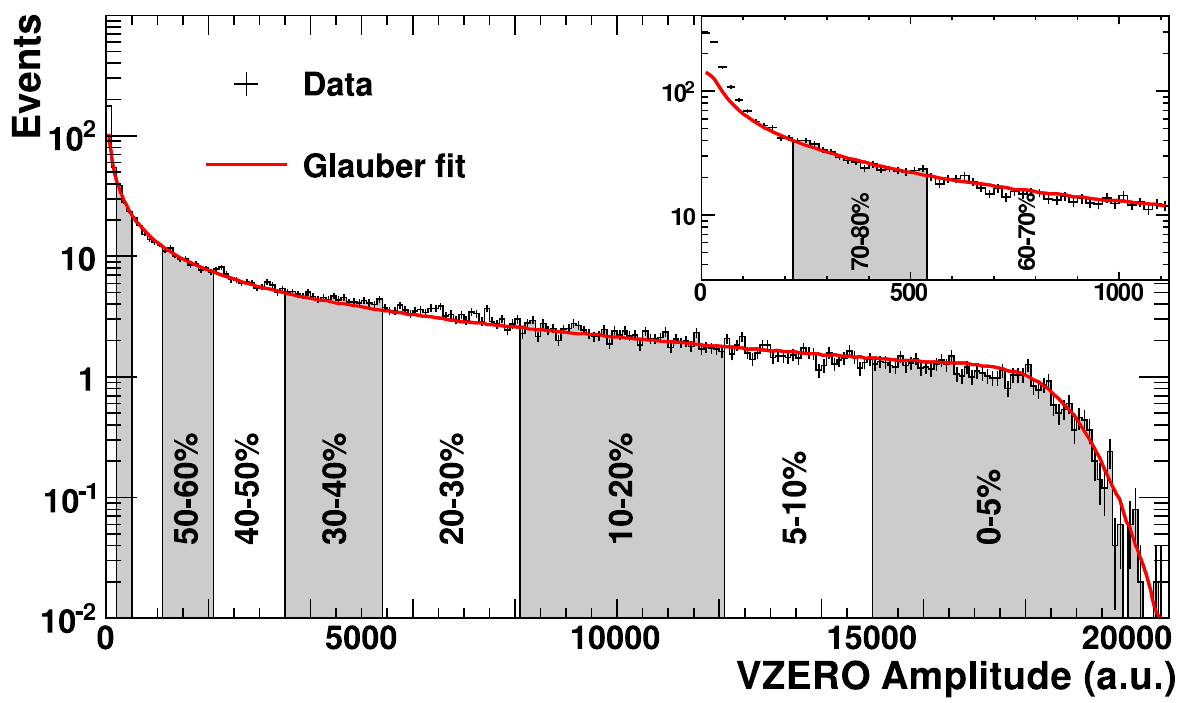}
\caption{\label{fig1}
Distribution of the summed amplitudes in the VZERO scintillator tiles (histogram); 
inset shows the low amplitude part of the distribution. 
The curve shows the result of the Glauber model fit to the measurement. 
The vertical lines separate the centrality classes used in the analysis, which in total
correspond to the most central 80\% of hadronic collisions.}
\end{center}
\end{figure}

After event selection, the sample consists of about \final{65\,000} events.
\Figure{fig1} shows the distribution of the summed amplitudes in the \VZERO\ scintillator tiles 
together with the distribution obtained with a model of particle production based on a Glauber 
description of nuclear collisions~\cite{Miller:2007ri}.
We use a two component model assuming that the number of particle-producing sources is given by 
$f\times\Npart + (1-f)\times\Ncoll$, where $\Npart$ is the number of participating nucleons, 
$\Ncoll$ is the number of binary nucleon--nucleon collisions and $f$ quantifies their relative contributions.
The number of particles produced by each source is distributed according to a Negative Binomial Distribution,
parametrized with $\mu$ and $\kappa$, where $\mu$ is the mean multiplicity 
per source and $\kappa$ controls the large multiplicity tail. 

In the Glauber calculation~\cite{Alver:2008aq}, the nuclear density for $^{208}$Pb is modeled by a Woods--Saxon
distribution for a spherical nucleus with a radius of $6.62$~fm and a skin depth of $0.546$~fm, based
on data from low energy electron--nucleus scattering experiments~\cite{DeJager:1987qc}.
A hard-sphere exclusion distance of $0.4$~fm between nucleons is employed.
Nuclear collisions are  modeled by randomly displacing the two colliding nuclei in the transverse plane.
Nucleons from each nucleus are assumed to collide if the transverse distance between 
them is less than the distance corresponding to the inelastic nucleon--nucleon cross section,
estimated from interpolating data at different centre-of-mass energies~\cite{Nakamura:2010zzi} 
to be $64\pm 5$~mb at $\sqrt{s} = 2.76$~TeV.

The values of $f$, $\mu$ and $\kappa$ are obtained from a fit to the measured \VZERO\ amplitude distribution.
The fit is restricted to amplitudes above a value corresponding to 88\% of the hadronic cross section.
In this region the trigger and event selection are fully efficient, and the contamination by electromagnetic 
processes is negligible. 
Centrality classes are determined by integrating the measured distribution above the cut, as shown in \Fig{fig1}.

The determination of $\dNdeta$ is performed for each centrality class.
The primary vertex position is extracted by correlating hits in the two SPD layers. 
All events in the sample corresponding to 0--80\% of the hadronic cross section are found to have a 
well-defined primary vertex. 
To minimize edge effects at the limit of the SPD acceptance, we require $|z_{\rm vtx}|<7$~cm
for the reconstructed vertex, leading to a sample of about \final{49\,000} events.

The measurement of the charged-particle multiplicity is based on the reconstruction of tracklets~\cite{Aamodt:2010pb}.
A tracklet candidate is defined as a pair of hits, one in each SPD layer.
Using the reconstructed vertex as the origin, differences in azimuthal~($\Delta\varphi$, bending plane) 
and polar~($\Delta\theta$, non-bending direction) angles for pairs of hits are calculated~\cite{alicepp2}. 
Tracklets are defined by hit combinations that satisfy a selection on the sum of the squares~($\delta^2$) of 
$\Delta\varphi$ and $\Delta\theta$, each normalized to its estimated resolution~(60~mrad for $\Delta\varphi$ 
and $25\,\rm{sin}^{2}\theta$~mrad for $\Delta\theta$).
The tolerance in $\Delta\varphi$ for tracklet reconstruction effectively selects charged particles with transverse 
momentum above 50~MeV/$c$. 
If multiple tracklet candidates share a hit, only the combination with the smallest $\delta^2$ is kept.

The charged-particle pseudo-rapidity density $\dNdeta$ in $\abs{\eta}<0.5 $ is obtained from the number of 
\mbox{tracklets} by applying a correction $\alpha\times(1-\beta)$ in bins of pseudo-rapidity and $z$-position of 
the primary vertex.
The factor $\alpha$ corrects for the acceptance and efficiency of a primary track to form a tracklet, 
and $\beta$ reflects the fraction of background tracklets from uncorrelated hits. 
The fraction $\beta$ is estimated by matching the tails of the data and background $\delta^{2}$ distributions.
The latter is obtained by selecting combinatorial tracklets from a sample of simulated events with similar 
SPD hit multiplicities generated with \mbox{HIJING}~\cite{hijing} and a \mbox{GEANT3}~\cite{geant3ref2} model 
of the detector response.
The estimated background fraction varies from \final{1}\% in the most peripheral to \final{14}\% in 
the most central class.

The correction $\alpha$ is obtained as the ratio of the number of generated primary charged particles and 
the number of reconstructed tracklets, after subtraction of the combinatorial background.
Thus, $\alpha$ includes the corrections for the geometrical acceptance, detector and reconstruction 
inefficiencies, contamination by weak decay products of strange particles, photon conversions, secondary 
interactions, and undetected particles with transverse momentum below 50 MeV/$c$.
The correction is about \final{1.8}  and varies little with centrality.
Its magnitude is dominated by the effect of tracklet acceptance: the fraction of SPD channels 
active during data taking was 70\% for the inner and 78\% for the outer layer.

{Systematic uncertainties} on $\dNdeta$ are estimated as follows:
for {\em background subtraction}, from 0.1\% in the most peripheral to 2.0\% in the most central class,
by using an alternative method where fake hits are injected into real events;
for {\em particle composition}, 1\%, by changing the relative abundances of protons, pions, kaons by up to a factor 
of two;
for {\em contamination by weak decays}, 1\%, by changing the relative contribution of the yield of strange particles
by a factor of two; 
for {\em extrapolation to zero transverse momentum}, 2\%, by varying the estimated yield of particles at low 
transverse momentum by a factor of two;
for dependence on {\em event generator}, 2\%, by using quenched and unquenched versions of HIJING~\cite{hijing}, 
as well as \mbox{DPMJET}~\cite{Roesler:2000he} for calculating the corrections.
The systematic uncertainty on $\dNdeta$ due to the {\em centrality class definition} is estimated as 6.2\% for the 
most peripheral and 0.4\% for the most central class,
by using alternative centrality definitions based on track or SPD hit multiplicities, 
by using different ranges for the Glauber model fit,
by defining cross-section classes integrating over the fit rather than directly over the data distributions,
by changing the $\Npart$ dependence of the particle production model to a power law,
and by changing the nucleon--nucleon cross section and the parameters of the Woods--Saxon distribution 
within their estimated uncertainties and by changing the inter-nucleon exclusion distance by $\pm100$\%.
All other sources of systematic errors considered~(tracklet cuts, vertex cuts, material budget, detector
efficiency, background events) were found to be negligible.
The total systematic uncertainty on $\dNdeta$ 
amounts to 7.0\% in the most peripheral and 3.8\% in the most central class.
A large part of this uncertainty, about 5.0\% for the most peripheral and 2.5\% for the most central class, 
is correlated among the different centrality classes.
The $\dNdeta$ values obtained for nine centrality classes together with their systematic uncertainties
are given in Table~\ref{tab1}.
As a cross-check of the centrality selection the $\dNdeta$ analysis was repeated using centrality cuts defined 
by slicing perpendicularly to the correlation between the energy deposited in the \ZDC\ and the \VZERO\ amplitude.
The resulting $\dNdeta$ values differ by 3.5\% in the most peripheral (70--80\%) and by less than 2\% in all the 
other classes from those obtained by using the \VZERO\ selection alone, which is well within
the systematic uncertainty.
Independent cross-checks performed using tracks reconstructed in the TPC and ITS instead of tracklets 
yield compatible results.

\begin{table}[tbhp]
\centering
\begin{tabular}{@{} c|c|c|c @{}} 
Centrality & $\dNdeta$ & $\avNpart$  & $\dNdetape$\\
\hline
0--5\%   &  $1601 \pm 60$  &  $382.8 \pm 3.1 $  &  $8.4 \pm 0.3 $\\
5--10\%  &  $1294 \pm 49$  &  $329.7 \pm 4.6 $  &  $7.9 \pm 0.3 $\\
10--20\% &  $966  \pm 37$  &  $260.5 \pm 4.4 $  &  $7.4 \pm 0.3 $\\
20--30\% &  $649  \pm 23$  &  $186.4 \pm 3.9 $  &  $7.0 \pm 0.3 $\\
30--40\% &  $426  \pm 15$  &  $128.9 \pm 3.3 $  &  $6.6 \pm 0.3 $\\
40--50\% &  $261  \pm 9 $  &  $85.0  \pm 2.6 $  &  $6.1 \pm 0.3 $\\
50--60\% &  $149  \pm 6 $  &  $52.8  \pm 2.0 $  &  $5.7 \pm 0.3 $\\
60--70\% &  $76   \pm 4 $  &  $30.0  \pm 1.3 $  &  $5.1 \pm 0.3 $\\
70--80\% &  $35   \pm 2 $  &  $15.8  \pm 0.6 $  &  $4.4 \pm 0.4 $\\
\end{tabular}
\caption{\label{tab1}
$\dNdeta$ and $\dNdetape$ values measured in $\abs{\eta}<0.5$ for nine centrality classes.
The $\avNpart$ obtained with the Glauber model are given.} 
\end{table}

\begin{figure}[tbh]
\begin{center}
\includegraphics[width=\figlen]{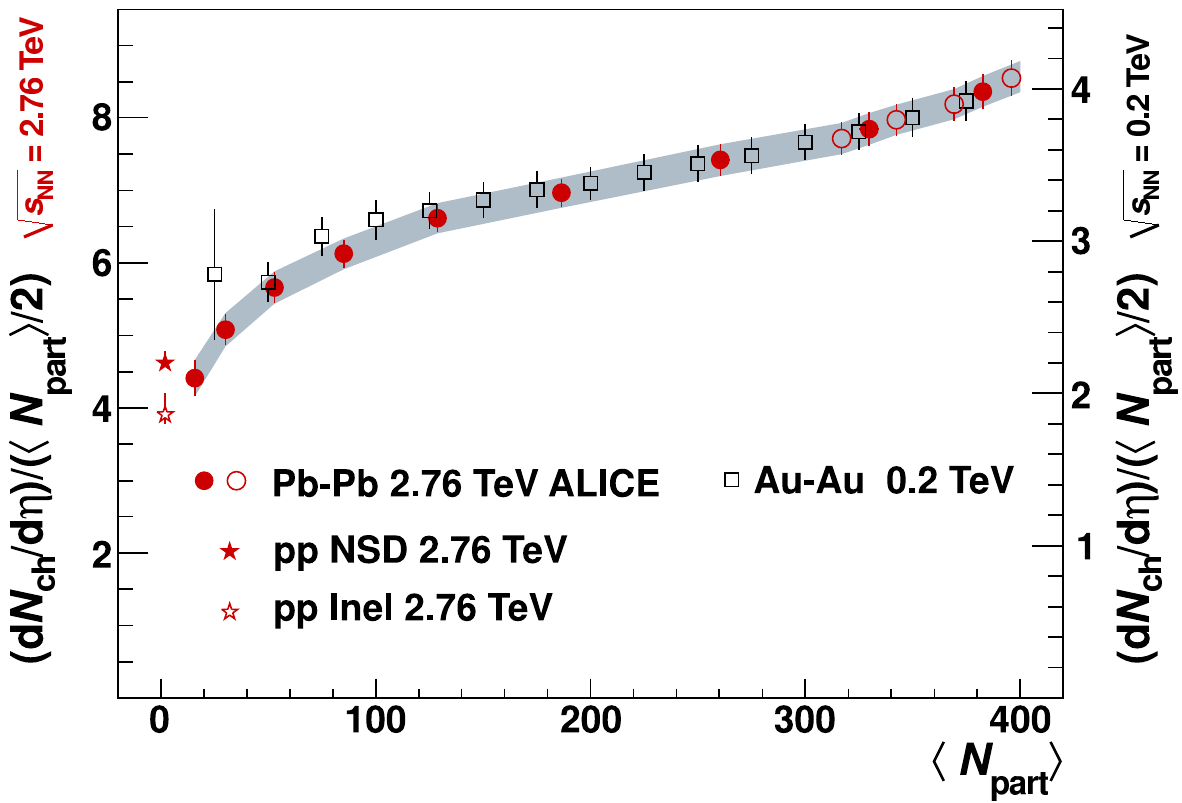}
\caption{\label{fig2}
Dependence of $\dNdetape$ on the number of participants for \PbPb\ collisions at $\snn=2.76$~TeV
and \AuAu\ collisions at $\snn=0.2$~TeV~(RHIC average)~\cite{Adler:2004zn}. 
The scale for the lower-energy data is shown on the right-hand side and differs from the scale for the 
higher-energy data on the left-hand side by a factor of 2.1.
For the \PbPb\ data, uncorrelated uncertainties are indicated by the error bars,
while correlated uncertainties are shown as the grey band.
Statistical errors are negligible. 
The open circles show the values obtained for centrality classes obtained by dividing the 0--10\%
most central collisions into four, rather than two classes.
The values for non-single-diffractive and inelastic \pp\ collisions are the results of interpolating between data
at 2.36~\cite{alicepp2,Khachatryan:2010xs} and 7 TeV~\cite{Khachatryan:2010us}.}
\end{center}
\end{figure}

In order to compare bulk particle production in different collision systems and at different energies, 
the charged-particle density is divided by the average number of participating nucleon pairs, 
$\avNpart/2$, determined for each centrality class.
The $\avNpart$ values are obtained using the Glauber calculation, by 
classifying events according to the impact parameter, without reference to a specific particle 
production model, and are listed in Table~\ref{tab1}.
The systematic uncertainty in the $\avNpart$ values is obtained by varying the parameters entering
the Glauber calculation as described above.
The geometrical $\avNpart$ values are consistent within uncertainties with the values extracted from the 
Glauber fit in each centrality class, and agree to better than 1\% except for the 70--80\% class 
where the difference is 3.5\%. 

Figure \ref{fig2} presents $\dNdetape$ as a function of the number of participants.
Point-to-point, uncorrelated uncertainties are indicated by the error bars,
while correlated uncertainties are shown as the grey band.
Statistical errors are negligible. 
The charged-particle density per participant pair increases with $\avNpart$, 
from $4.4\pm0.4$ for the most peripheral to $8.4\pm0.3$ for the most central class.
The values for \AuAu\ collisions at $\snn=0.2$~TeV, averaged over the RHIC experiments~\cite{Adler:2004zn},
are shown in the same figure with a scale that differs by a factor of 2.1 on the right-hand side.
The centrality dependence of the multiplicity is found to be very similar for 
$\snn = 2.76$~TeV and $\snn=0.2$~TeV.

\begin{figure}[tbh]
\begin{center}
\includegraphics[width=\figlen]{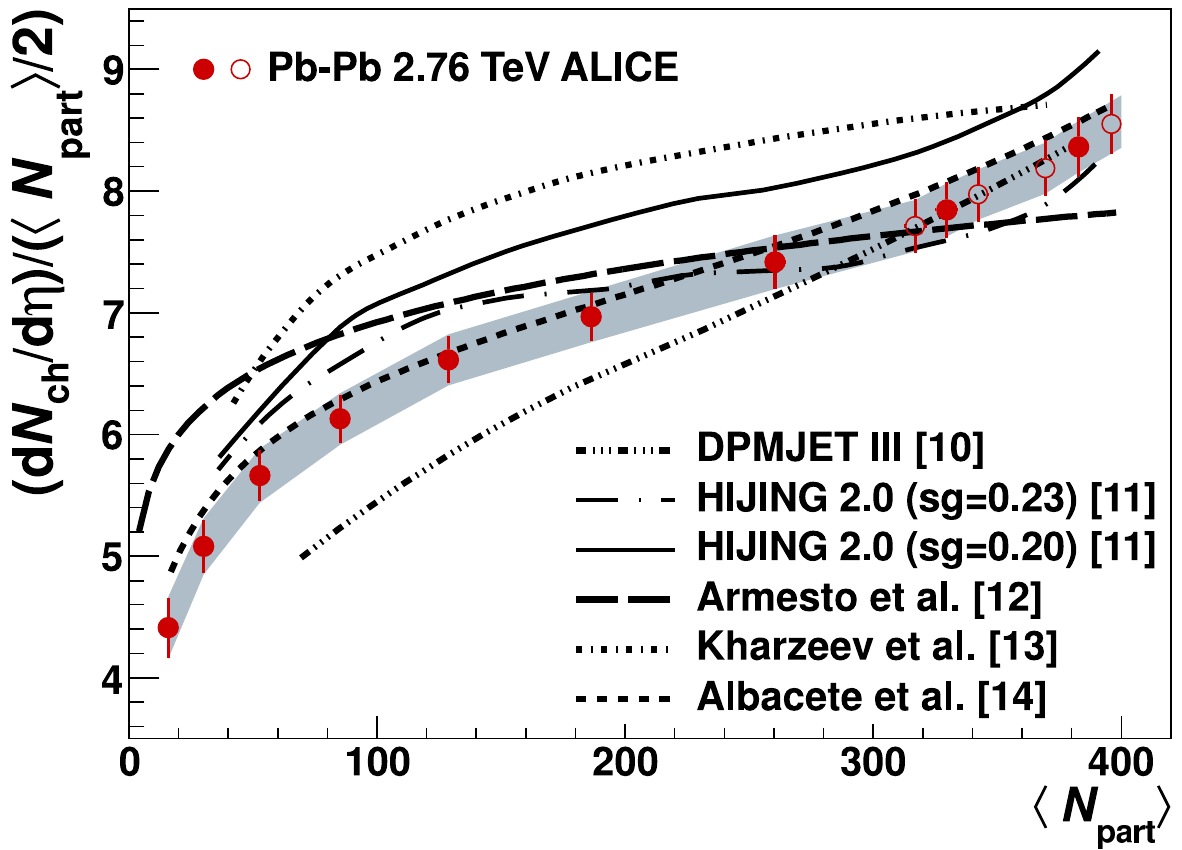}
\caption{\label{fig3}
Comparison of $\dNdetape$ with model calculations for \PbPb\ at $\snn=2.76$~TeV. 
Uncertainties in the data are shown as in \Fig{fig2}.
The HIJING 2.0 curve is shown for two values of the gluon shadowing~($s_{g}$) parameter.}
\end{center}
\end{figure}

Theoretical descriptions of particle production in nuclear collisions fall into two broad categories: 
two-component models combining perturbative QCD processes (e.g.\ jets and mini-jets) with soft 
interactions, and saturation models with various parametrizations for the energy and centrality 
dependence of the saturation scale.
In \Fig{fig3} we compare the measured $\dNdetape$ with model predictions.
A calculation based on the two-component Dual Parton Model~(DPMJET~\cite{Bopp:2007sa}, with string fusion)
exhibits a stronger rise with centrality than observed.
The two-component HIJING 2.0 model~\cite{Deng:2010mv}, which has been tuned~\cite{Deng:2010xg}\footnote{Published 
after the most central $\dNdeta$ value~\cite{Aamodt:2010pb} was known.} 
to high-energy \pp~\cite{alicepp2,Khachatryan:2010xs} and central \PbPb\ data~\cite{Aamodt:2010pb}, 
reasonably describes the data. 
This model includes a strong impact parameter dependent gluon shadowing~($s_{g}$) 
which limits the rise of particle production with centrality. 
The remaining models show a weak dependence of multiplicity on centrality.
They are all different implementations of the saturation picture, 
where the number of soft gluons available for scattering and particle production is 
reduced by nonlinear interactions and parton recombination.
A geometrical scaling model with a strong dependence of the saturation scale on nuclear mass and collision
energy~\cite{Armesto:2004ud} predicts a rather weak variation with centrality.
The centrality dependence is well reproduced by saturation models~\cite{Kharzeev:2004if} and 
\cite{ALbacete:2010ad}\footnotemark[1], although the former overpredicts the magnitude.

In summary, the measurement of the centrality dependence of the charged-particle multiplicity density at mid-rapidity
in \PbPb\ collisions at $\snn=2.76$ TeV has been presented.
The charged-particle density normalized per participating nucleon pair increases by about a factor 2 from 
peripheral~(70--80\%) to central~(0--5\%) collisions. 
The dependence of the multiplicity on centrality is strikingly similar for the data at $\snn = 2.76$~TeV 
and $\snn=0.2$~TeV.
Theoretical descriptions that include a moderation of the multiplicity evolution with centrality are 
favoured by the data.

\ifpreprint
\newenvironment{acknowledgement}{\relax}{\relax}
\begin{acknowledgement}
\section*{Acknowledgements}
The ALICE collaboration would like to thank all its engineers and technicians for their invaluable contributions to the construction of the experiment and the CERN accelerator teams for the outstanding performance of the LHC complex.
The ALICE collaboration acknowledges the following funding agencies for their support in building and
running the ALICE detector:
Calouste Gulbenkian Foundation from Lisbon and Swiss Fonds Kidagan, Armenia;
Conselho Nacional de Desenvolvimento Cient\'{\i}fico e Tecnol\'{o}gico (CNPq), Financiadora de Estudos e Projetos (FINEP),
Funda\c{c}\~{a}o de Amparo \`{a} Pesquisa do Estado de S\~{a}o Paulo (FAPESP);
National Natural Science Foundation of China (NSFC), the Chinese Ministry of Education (CMOE)
and the Ministry of Science and Technology of China (MSTC);
Ministry of Education and Youth of the Czech Republic;
Danish Natural Science Research Council, the Carlsberg Foundation and the Danish National Research Foundation;
The European Research Council under the European Community's Seventh Framework Programme;
Helsinki Institute of Physics and the Academy of Finland;
French CNRS-IN2P3, the `Region Pays de Loire', `Region Alsace', `Region Auvergne' and CEA, France;
German BMBF and the Helmholtz Association;
Greek Ministry of Research and Technology;
Hungarian OTKA and National Office for Research and Technology (NKTH);
Department of Atomic Energy and Department of Science and Technology of the Government of India;
Istituto Nazionale di Fisica Nucleare (INFN) of Italy;
MEXT Grant-in-Aid for Specially Promoted Research, Ja\-pan;
Joint Institute for Nuclear Research, Dubna;
 %
National Research Foundation of Korea (NRF);
CONACYT, DGAPA, M\'{e}xico, ALFA-EC and the HELEN Program (High-Energy physics Latin-American--European Network);
Stichting voor Fundamenteel Onderzoek der Materie (FOM) and the Nederlandse Organisatie voor Wetenschappelijk Onderzoek (NWO), Netherlands;
Research Council of Norway (NFR);
Polish Ministry of Science and Higher Education;
National Authority for Scientific Research - NASR (Autoritatea Na\c{t}ional\u{a} pentru Cercetare \c{S}tiin\c{t}ific\u{a} - ANCS);
Federal Agency of Science of the Ministry of Education and Science of Russian Federation, International Science and
Technology Center, Russian Academy of Sciences, Russian Federal Agency of Atomic Energy, Russian Federal Agency for Science and Innovations and CERN-INTAS;
Ministry of Education of Slovakia;
CIEMAT, EELA, Ministerio de Educaci\'{o}n y Ciencia of Spain, Xunta de Galicia (Conseller\'{\i}a de Educaci\'{o}n),
CEA\-DEN, Cubaenerg\'{\i}a, Cuba, and IAEA (International Atomic Energy Agency);
The Ministry of Science and Technology and the National Research Foundation (NRF), South Africa;
Swedish Reseach Council (VR) and Knut $\&$ Alice Wallenberg Foundation (KAW);
Ukraine Ministry of Education and Science;
United Kingdom Science and Technology Facilities Council (STFC);
The United States Department of Energy, the United States National
Science Foundation, the State of Texas, and the State of Ohio.
\end{acknowledgement}
\ifbibtex
\bibliographystyle{plain}
\bibliography{multPbPb}{}
\else

\fi
\newpage
\appendix
\section{The ALICE Collaboration}
\label{app:collab}
%
\begingroup
\small
\begin{flushleft}
K.~Aamodt\Irefn{0}\And
A.~Abrahantes~Quintana\Irefn{1}\And
D.~Adamov\'{a}\Irefn{2}\And
A.M.~Adare\Irefn{3}\And
M.M.~Aggarwal\Irefn{4}\And
G.~Aglieri~Rinella\Irefn{5}\And
A.G.~Agocs\Irefn{6}\And
S.~Aguilar~Salazar\Irefn{7}\And
Z.~Ahammed\Irefn{8}\And
N.~Ahmad\Irefn{9}
% ...
\renewcommand\labelenumi{\textsuperscript{\theenumi}~}
\section*{Affiliation notes}
\renewcommand\theenumi{\roman{enumi}}
\begin{Authlist}
\item \Adef{0}Also at Laboratoire de Physique Corpusculaire (LPC), Clermont Universit\'{e}, Universit\'{e} Blaise Pascal, CNRS--IN2P3, Clermont-Ferrand, France
\item \Adef{1}Now at Centro Fermi -- Centro Studi e Ricerche e Museo Storico della Fisica ``Enrico Fermi'', Rome, Italy
\item \Adef{2}Now at European Organization for Nuclear Research (CERN), Geneva, Switzerland
\item \Adef{3}Also at European Organization for Nuclear Research (CERN), Geneva, Switzerland
\item \Adef{4}Now at Physikalisches Institut, Ruprecht-Karls-Universit\"{a}t Heidelberg, Heidelberg, Germany
\item \Adef{5}Now at Sezione INFN, Turin, Italy
\item \Adef{6}Now at University of Houston, Houston, Texas, United States
\item \Adef{7}Also at Dipartimento di Fisica Sperimentale dell'Universit\`{a} and Sezione INFN, Turin, Italy
\item \Adef{8}Also at  Dipartimento di Fisica dell'Universit\'{a}, Udine, Italy 
\item \Adef{9}Now at SUBATECH, Ecole des Mines de Nantes, Universit\'{e} de Nantes, CNRS-IN2P3, Nantes, France
\item \Adef{10}Now at Centro de Investigaci\'{o}n y de Estudios Avanzados (CINVESTAV), Mexico City and M\'{e}rida, Mexico
\item \Adef{11}Now at Benem\'{e}rita Universidad Aut\'{o}noma de Puebla, Puebla, Mexico
\item \Adef{12}Now at Laboratoire de Physique Subatomique et de Cosmologie (LPSC), Universit\'{e} Joseph Fourier, CNRS-IN2P3, Institut Polytechnique de Grenoble, Grenoble, France
\item \Adef{13}Now at Institut Pluridisciplinaire Hubert Curien (IPHC), Universit\'{e} de Strasbourg, CNRS-IN2P3, Strasbourg, France
\item \Adef{14}Now at Sezione INFN, Padova, Italy
\item \Adef{15} Deceased 
\item \Adef{16}Also at Division of Experimental High Energy Physics, University of Lund, Lund, Sweden
\item \Adef{17}Also at  University of Technology and Austrian Academy of Sciences, Vienna, Austria 
\item \Adef{18}Now at Oak Ridge National Laboratory, Oak Ridge, Tennessee, United States
\item \Adef{19}Also at Wayne State University, Detroit, Michigan, United States
\item \Adef{20}Also at Frankfurt Institute for Advanced Studies, Johann Wolfgang Goethe-Universit\"{a}t Frankfurt, Frankfurt, Germany
\item \Adef{21}Now at Frankfurt Institute for Advanced Studies, Johann Wolfgang Goethe-Universit\"{a}t Frankfurt, Frankfurt, Germany
\item \Adef{22}Now at Research Division and ExtreMe Matter Institute EMMI, GSI Helmholtzzentrum f\"ur Schwerionenforschung, Darmstadt, Germany
\item \Adef{23}Also at Fachhochschule K\"{o}ln, K\"{o}ln, Germany
\item \Adef{24}Also at Institute of Experimental Physics, Slovak Academy of Sciences, Ko\v{s}ice, Slovakia
\item \Adef{25}Now at Instituto de Ciencias Nucleares, Universidad Nacional Aut\'{o}noma de M\'{e}xico, Mexico City, Mexico
\item \Adef{26}Also at Laboratoire de Physique Subatomique et de Cosmologie (LPSC), Universit\'{e} Joseph Fourier, CNRS-IN2P3, Institut Polytechnique de Grenoble, Grenoble, France
\item \Adef{27}Also at  "Vin\v{c}a" Institute of Nuclear Sciences, Belgrade, Serbia 
\item \Adef{28}Also at Dipartimento di Fisica `E.R.~Caianiello' dell'Universit\`{a} and Gruppo Collegato INFN, Salerno, Italy
\item \Adef{29}Also at Instituto de Ciencias Nucleares, Universidad Nacional Aut\'{o}noma de M\'{e}xico, Mexico City, Mexico
\item \Adef{30}Also at University of Houston, Houston, Texas, United States
\item \Adef{31}Also at Department of Physics, University of Oslo, Oslo, Norway
\item \Adef{32}Now at Department of Physics, University of Oslo, Oslo, Norway
\item \Adef{33}Also at Eberhard Karls Universit\"{a}t T\"{u}bingen, T\"{u}bingen, Germany
\item \Adef{34}Also at Dipartimento Interateneo di Fisica `M.~Merlin' and Sezione INFN, Bari, Italy
\item \Adef{35}Now at Nikhef, National Institute for Subatomic Physics and Institute for Subatomic Physics of Utrecht University, Utrecht, Netherlands
\item \Adef{36}Also at Hua-Zhong Normal University, Wuhan, China
\item \Adef{37}Also at Centro Fermi -- Centro Studi e Ricerche e Museo Storico della Fisica ``Enrico Fermi'', Rome, Italy
\end{Authlist}
\section*{Collaboration Institutes}
\renewcommand\theenumi{\arabic{enumi}~}
\begin{Authlist}
\item \Idef{0}Department of Physics and Technology, University of Bergen, Bergen, Norway
\item \Idef{1}Centro de Aplicaciones Tecnol\'{o}gicas y Desarrollo Nuclear (CEADEN), Havana, Cuba
\item \Idef{2}Nuclear Physics Institute, Academy of Sciences of the Czech Republic, \v{R}e\v{z} u Prahy, Czech Republic
\item \Idef{3}Yale University, New Haven, Connecticut, United States
\item \Idef{4}Physics Department, Panjab University, Chandigarh, India
\item \Idef{5}European Organization for Nuclear Research (CERN), Geneva, Switzerland
\item \Idef{6}KFKI Research Institute for Particle and Nuclear Physics, Hungarian Academy of Sciences, Budapest, Hungary
\item \Idef{7}Instituto de F\'{\i}sica, Universidad Nacional Aut\'{o}noma de M\'{e}xico, Mexico City, Mexico
\item \Idef{8}Variable Energy Cyclotron Centre, Kolkata, India
\item \Idef{9}Department of Physics Aligarh Muslim University, Aligarh, India
\item \Idef{10}Gangneung-Wonju National University, Gangneung, South Korea
\item \Idef{11}Institute for Theoretical and Experimental Physics, Moscow, Russia
\item \Idef{12}Russian Research Centre Kurchatov Institute, Moscow, Russia
\item \Idef{13}Sezione INFN, Turin, Italy
\item \Idef{14}Dipartimento di Fisica dell'Universit\`{a} and Sezione INFN, Bologna, Italy
\item \Idef{15}Bogolyubov Institute for Theoretical Physics, Kiev, Ukraine
\item \Idef{16}Frankfurt Institute for Advanced Studies, Johann Wolfgang Goethe-Universit\"{a}t Frankfurt, Frankfurt, Germany
\item \Idef{17}Dipartimento Interateneo di Fisica `M.~Merlin' and Sezione INFN, Bari, Italy
\item \Idef{18}Research Division and ExtreMe Matter Institute EMMI, GSI Helmholtzzentrum f\"ur Schwerionenforschung, Darmstadt, Germany
\item \Idef{19}V.~Fock Institute for Physics, St. Petersburg State University, St. Petersburg, Russia
\item \Idef{20}National Institute for Physics and Nuclear Engineering, Bucharest, Romania
\item \Idef{21}Kirchhoff-Institut f\"{u}r Physik, Ruprecht-Karls-Universit\"{a}t Heidelberg, Heidelberg, Germany
\item \Idef{22}Department of Physics, Ohio State University, Columbus, Ohio, United States
\item \Idef{23}Rudjer Bo\v{s}kovi\'{c} Institute, Zagreb, Croatia
\item \Idef{24}Sezione INFN, Padova, Italy
\item \Idef{25}Sezione INFN, Bologna, Italy
\item \Idef{26}SUBATECH, Ecole des Mines de Nantes, Universit\'{e} de Nantes, CNRS-IN2P3, Nantes, France
\item \Idef{27}Institut f\"{u}r Kernphysik, Johann Wolfgang Goethe-Universit\"{a}t Frankfurt, Frankfurt, Germany
\item \Idef{28}Laboratoire de Physique Subatomique et de Cosmologie (LPSC), Universit\'{e} Joseph Fourier, CNRS-IN2P3, Institut Polytechnique de Grenoble, Grenoble, France
\item \Idef{29}Departamento de F\'{\i}sica de Part\'{\i}culas and IGFAE, Universidad de Santiago de Compostela, Santiago de Compostela, Spain
\item \Idef{30}Oak Ridge National Laboratory, Oak Ridge, Tennessee, United States
\item \Idef{31}Helsinki Institute of Physics (HIP) and University of Jyv\"{a}skyl\"{a}, Jyv\"{a}skyl\"{a}, Finland
\item \Idef{32}Sezione INFN, Catania, Italy
\item \Idef{33}Dipartimento di Fisica Sperimentale dell'Universit\`{a} and Sezione INFN, Turin, Italy
\item \Idef{34}Centro Fermi -- Centro Studi e Ricerche e Museo Storico della Fisica ``Enrico Fermi'', Rome, Italy
\item \Idef{35}Commissariat \`{a} l'Energie Atomique, IRFU, Saclay, France
\item \Idef{36}Laboratoire de Physique Corpusculaire (LPC), Clermont Universit\'{e}, Universit\'{e} Blaise Pascal, CNRS--IN2P3, Clermont-Ferrand, France
\item \Idef{37}Institute of Experimental Physics, Slovak Academy of Sciences, Ko\v{s}ice, Slovakia
\item \Idef{38}Dipartimento di Fisica e Astronomia dell'Universit\`{a} and Sezione INFN, Catania, Italy
\item \Idef{39}School of Physics and Astronomy, University of Birmingham, Birmingham, United Kingdom
\item \Idef{40}The Henryk Niewodniczanski Institute of Nuclear Physics, Polish Academy of Sciences, Cracow, Poland
\item \Idef{41}Institut f\"{u}r Kernphysik, Westf\"{a}lische Wilhelms-Universit\"{a}t M\"{u}nster, M\"{u}nster, Germany
\item \Idef{42}Joint Institute for Nuclear Research (JINR), Dubna, Russia
\item \Idef{43}Niels Bohr Institute, University of Copenhagen, Copenhagen, Denmark
\item \Idef{44}Institut Pluridisciplinaire Hubert Curien (IPHC), Universit\'{e} de Strasbourg, CNRS-IN2P3, Strasbourg, France
\item \Idef{45}Wayne State University, Detroit, Michigan, United States
\item \Idef{46}Petersburg Nuclear Physics Institute, Gatchina, Russia
\item \Idef{47}Physics Department, University of Jammu, Jammu, India
\item \Idef{48}Laboratori Nazionali di Frascati, INFN, Frascati, Italy
\item \Idef{49}Dipartimento di Fisica dell'Universit\`{a} and Sezione INFN, Padova, Italy
\item \Idef{50}Faculty of Nuclear Sciences and Physical Engineering, Czech Technical University in Prague, Prague, Czech Republic
\item \Idef{51}Nikhef, National Institute for Subatomic Physics, Amsterdam, Netherlands
\item \Idef{52}Centro de Investigaciones Energ\'{e}ticas Medioambientales y Tecnol\'{o}gicas (CIEMAT), Madrid, Spain
\item \Idef{53}University of Houston, Houston, Texas, United States
\item \Idef{54}Moscow Engineering Physics Institute, Moscow, Russia
\item \Idef{55}Institute for High Energy Physics, Protvino, Russia
\item \Idef{56}Faculty of Science, P.J.~\v{S}af\'{a}rik University, Ko\v{s}ice, Slovakia
\item \Idef{57}Saha Institute of Nuclear Physics, Kolkata, India
\item \Idef{58}Institut de Physique Nucl\'{e}aire d'Orsay (IPNO), Universit\'{e} Paris-Sud, CNRS-IN2P3, Orsay, France
\item \Idef{59}Department of Physics, University of Oslo, Oslo, Norway
\item \Idef{60}Dipartimento di Fisica dell'Universit\`{a} and Sezione INFN, Trieste, Italy
\item \Idef{61}Faculty of Mathematics, Physics and Informatics, Comenius University, Bratislava, Slovakia
\item \Idef{62}Russian Federal Nuclear Center (VNIIEF), Sarov, Russia
\item \Idef{63}Physikalisches Institut, Ruprecht-Karls-Universit\"{a}t Heidelberg, Heidelberg, Germany
\item \Idef{64}Physics Department, University of Cape Town, iThemba LABS, Cape Town, South Africa
\item \Idef{65}Hua-Zhong Normal University, Wuhan, China
\item \Idef{66}Secci\'{o}n F\'{\i}sica, Departamento de Ciencias, Pontificia Universidad Cat\'{o}lica del Per\'{u}, Lima, Peru
\item \Idef{67}Physics Department, Creighton University, Omaha, Nebraska, United States
\item \Idef{68}Universit\'{e} de Lyon, Universit\'{e} Lyon 1, CNRS/IN2P3, IPN-Lyon, Villeurbanne, France
\item \Idef{69}Universidade Estadual de Campinas (UNICAMP), Campinas, Brazil
\item \Idef{70}Nikhef, National Institute for Subatomic Physics and Institute for Subatomic Physics of Utrecht University, Utrecht, Netherlands
\item \Idef{71}Division of Experimental High Energy Physics, University of Lund, Lund, Sweden
\item \Idef{72}University of Tsukuba, Tsukuba, Japan
\item \Idef{73}Sezione INFN, Cagliari, Italy
\item \Idef{74}Centro de Investigaci\'{o}n y de Estudios Avanzados (CINVESTAV), Mexico City and M\'{e}rida, Mexico
\item \Idef{75}Benem\'{e}rita Universidad Aut\'{o}noma de Puebla, Puebla, Mexico
\item \Idef{76}Dipartimento di Scienze e Tecnologie Avanzate dell'Universit\`{a} del Piemonte Orientale and Gruppo Collegato INFN, Alessandria, Italy
\item \Idef{77}Instituto de Ciencias Nucleares, Universidad Nacional Aut\'{o}noma de M\'{e}xico, Mexico City, Mexico
\item \Idef{78}Laboratori Nazionali di Legnaro, INFN, Legnaro, Italy
\item \Idef{79}Institute of Space Sciences (ISS), Bucharest, Romania
\item \Idef{80}Institute of Physics, Bhubaneswar, India
\item \Idef{81}Universidade de S\~{a}o Paulo (USP), S\~{a}o Paulo, Brazil
\item \Idef{82}Dipartimento di Fisica `E.R.~Caianiello' dell'Universit\`{a} and Gruppo Collegato INFN, Salerno, Italy
\item \Idef{83}Sezione INFN, Bari, Italy
\item \Idef{84}Dipartimento di Fisica dell'Universit\`{a} and Sezione INFN, Cagliari, Italy
\item \Idef{85}Soltan Institute for Nuclear Studies, Warsaw, Poland
\item \Idef{86}Sezione INFN, Rome, Italy
\item \Idef{87}Faculty of Engineering, Bergen University College, Bergen, Norway
\item \Idef{88}Institute for Nuclear Research, Academy of Sciences, Moscow, Russia
\item \Idef{89}Sezione INFN, Trieste, Italy
\item \Idef{90}Physics Department, University of Athens, Athens, Greece
\item \Idef{91}Warsaw University of Technology, Warsaw, Poland
\item \Idef{92}Universidad Aut\'{o}noma de Sinaloa, Culiac\'{a}n, Mexico
\item \Idef{93}Technical University of Split FESB, Split, Croatia
\item \Idef{94}Yerevan Physics Institute, Yerevan, Armenia
\item \Idef{95}University of Tokyo, Tokyo, Japan
\item \Idef{96}Department of Physics, Sejong University, Seoul, South Korea
\item \Idef{97}Lawrence Berkeley National Laboratory, Berkeley, California, United States
\item \Idef{98}Indian Institute of Technology, Mumbai, India
\item \Idef{99}Institut f\"{u}r Kernphysik, Technische Universit\"{a}t Darmstadt, Darmstadt, Germany
\item \Idef{100}Yonsei University, Seoul, South Korea
\item \Idef{101}Zentrum f\"{u}r Technologietransfer und Telekommunikation (ZTT), Fachhochschule Worms, Worms, Germany
\item \Idef{102}California Polytechnic State University, San Luis Obispo, California, United States
\item \Idef{103}China Institute of Atomic Energy, Beijing, China
\item \Idef{104}Institute of Physics, Academy of Sciences of the Czech Republic, Prague, Czech Republic
\item \Idef{105}University of Tennessee, Knoxville, Tennessee, United States
\item \Idef{106}Dipartimento di Fisica dell'Universit\`{a} `La Sapienza' and Sezione INFN, Rome, Italy
\item \Idef{107}Hiroshima University, Hiroshima, Japan
\item \Idef{108}Lawrence Livermore National Laboratory, Livermore, California, United States
\item \Idef{109}Budker Institute for Nuclear Physics, Novosibirsk, Russia
\item \Idef{110}Physics Department, University of Rajasthan, Jaipur, India
\item \Idef{111}Purdue University, West Lafayette, Indiana, United States
\item \Idef{112}Centre de Calcul de l'IN2P3, Villeurbanne, France 
\item \Idef{113}Pusan National University, Pusan, South Korea
\end{Authlist}
\endgroup
\else
\iffull

\else
\ifbibtex
\bibliographystyle{apsrev4-1}
\bibliography{multPbPb}{}
\else

\fi
\fi
\fi
\end{document}